\documentstyle[12pt,clustering]{article}

\input epsf.sty

\begin{document}
\newcommand{\sP}{\mbox{\protect\boldmath $P$}}
\newcommand{\sxi}{\mbox{\protect\boldmath $\xi$}}
\newcommand{\sk}{\mbox{\protect\boldmath $k$}}
\newcommand{\ff}{f}
\newcommand{\kms}{{\rm km}\,{\rm s}^{-1}}
\heading{REDSHIFT DISTORTIONS AND OMEGA\\
         IN IRAS SURVEYS}
\author{A.\ J.\ S.\ Hamilton}
       {JILA, U.\ Colorado, Boulder, CO 80309, USA.}

\begin{abstract}{\baselineskip 0.4cm 
Redshift space distortions on large scales can be used to measure
the linear growth rate parameter $\ff \approx \Omega^{0.6}/b$.
I report here measurements of such distortions in the
{\em IRAS} 2~Jy, 1.2~Jy, and QDOT redshift surveys,
finding $\ff = 0.69^{+ .21}_{- .19}$ from a merged QDOT plus 1.2~Jy catalogue.
Unfortunately, confidence in this result is undermined by a marked ($4\sigma$)
change in the pattern of clustering in QDOT beyond about $80 h^{-1} {\rm Mpc}$.
A similar effect may be present at a mild level in the 1.2~Jy survey.
The effect may be caused by systematic variation in the effective flux
limit of the {\em IRAS} PSC over the sky,
with a dispersion of $\sim 0.1$~Jy on scales $\sim 7^{\circ}$.
If so, then the value of $\ff$ inferred from redshift distortions
in {\em IRAS\/} surveys may be systematically underestimated.
}
\end{abstract}

\section{Redshift Correlation Function}
Peculiar velocities cause line of sight distortions in the pattern
of clustering of galaxies observed in redshift space.
The most familiar distortion is the finger-of-god effect,
caused by large velocities of galaxies in collapsed clusters.
A more subtle effect is the squashing effect on large scales
induced by peculiar infall towards overdense regions.
Both kinds of distortion can be used to estimate the cosmological density
$\Omega$, from the virial theorem on small scales,
and from linear theory on large scales.

\begin{figure}
\begin{center}
\leavevmode
\epsfbox{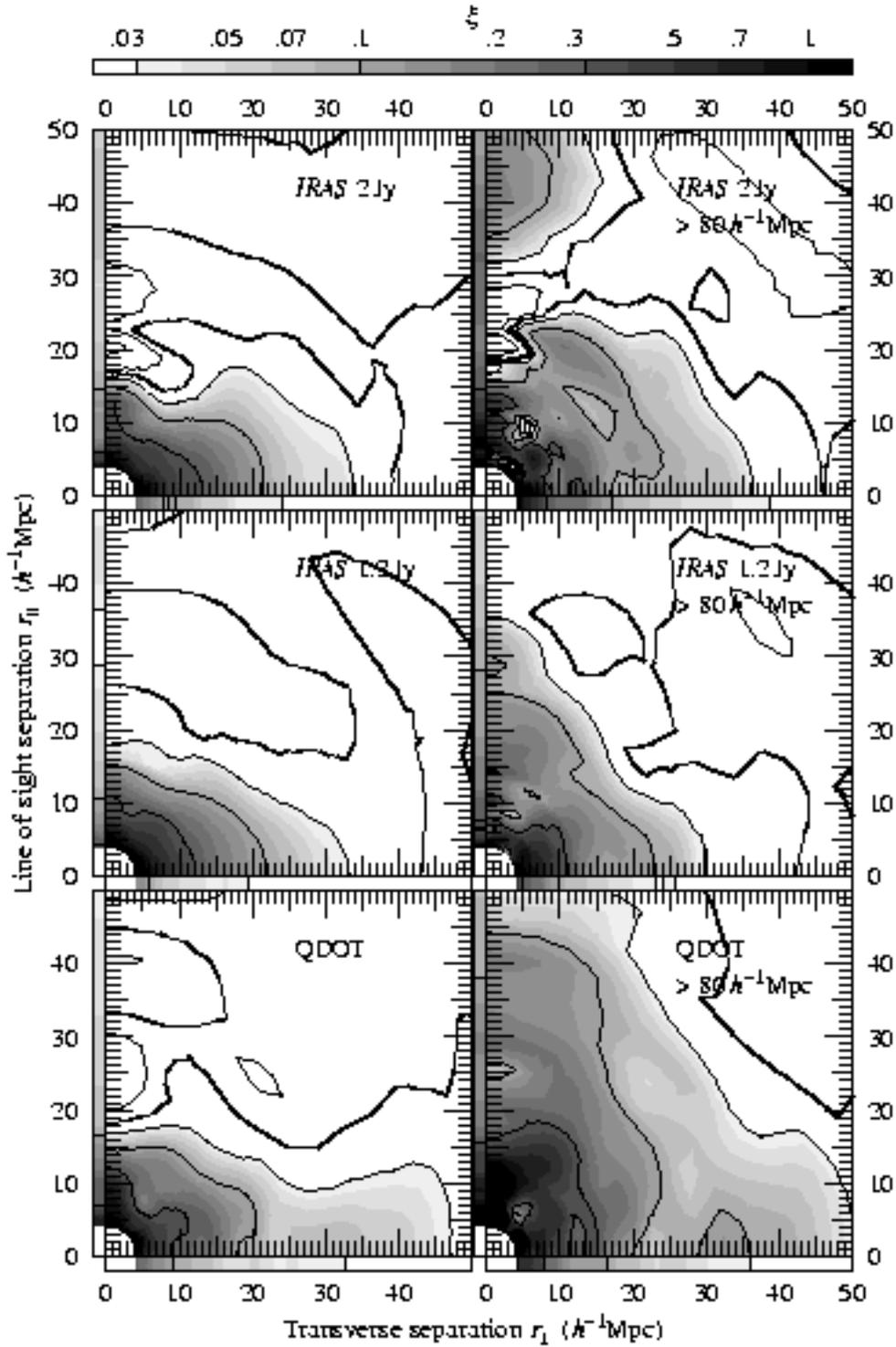}
\end{center}
\caption{
\label{xicont}
\baselineskip 0.4cm
\protect\small
Contour plots of the redshift space two-point correlation function $\sxi$
in the {\em IRAS} 2~Jy (top), 1.2~Jy (middle), and QDOT (bottom) surveys.
Left panels show the full survey in each case
(excluding the region within $25 h^{-1} {\rm Mpc}$
so as to eliminate local bias),
while right panels show the region beyond $80 h^{-1} {\rm Mpc}$.
The thick contour signifies $\sxi = 0$,
and other contours are logarithmically spaced
at intervals of 0.5~dex above $10^{-1.5}$,
or below $-10^{-1.5}$ for negative contours.
Shading is graduated at intervals of 0.1~dex above $10^{-1.5}$.
The correlation function here has been smoothed over pair separation
$r = ( r_\perp^2 + r_\parallel^2 )^{1/2}$
with a tophat window of width $0.2$~dex,
and over angles $\theta = \tan^{-1} ( r_\perp / r_\parallel )$
to the line of sight with a gaussian window
with a 1$\sigma$ width of $10^\circ$.
Bars along the two axes of each frame are contour plots of the 1$\sigma$
uncertainties in the plotted correlation function in the two particular
directions parallel ($\theta = 0^\circ$)
and perpendicular ($\theta = 90^\circ$)
to the line of sight.
The uncertainties are greater along the line of sight axis
because the volume element $\sin \theta d \theta$ is smaller.
}
\end{figure}

The recent growing interest in the large scale squashing effect
has been encouraged by the availability of the {\em IRAS\/}
redshift surveys,
which seem for the first time to probe a sufficiently large
volume of the Universe to allow a statistically significant
measurement of $\Omega$ from the effect.
Figure~\ref{xicont} illustrates the distortion of the redshift space
correlation function $\sxi$ in
the {\em IRAS} 2~Jy
\cite{S90}, \cite{S92a}, \cite{S92b},
1.2~Jy
\cite{12Jy},
and revised QDOT
\cite{Lawrence}
redshift surveys.
In each case, the full survey (left panels)
shows plainly the expected large scale squashing effect,
while fingers-of-god show up as the
$\sim 4 h^{-1} {\rm Mpc}$ wide by $15 h^{-1} {\rm Mpc}$ long
enhancements along the line of sight axis.

\section{Theory}
The mathematical relation between $\Omega$ and the squashing effect
was clarified by Kaiser \cite{Kaiser}
whose approach demonstrated, amongst other things,
the power of working in Fourier space when in the linear regime.
Kaiser showed that in the linear regime
a wave of amplitude $\delta ({\sk})$ appears amplified in redshift
space by a factor $1 + \ff \mu^2$,
where $\mu$ is the cosine of the angle between
the wavevector ${\sk}$ and the line of sight,
and $\ff$ is the growth rate of growing modes in linear theory,
which is related to $\Omega$ by
\begin{equation}
\label{fb}
 \ff \approx {\Omega^{0.6} \over b}
\end{equation}
in standard pressureless Friedmann cosmology with mass-to-light bias $b$.
Kaiser concluded that the redshift space power spectrum
$\sP ({\sk})$,
the Fourier transform of the correlation function $\sxi$,
appears amplified
by a factor $( 1 + \ff \mu^2 )^2$ over the true power spectrum
$P ( k )$
(boldface distinguishes redshift space power spectra and correlation functions
from their unredshifted counterparts)
\begin{equation}
\label{Kaiser}
 \sP ( {\sk} ) =
  ( 1 + \ff \mu^2 )^2 P ( k ) \ .
\end{equation}

Equation (\ref{Kaiser}) predicts that the redshift space power spectrum
in the linear regime is a sum of monopole, quadrupole,
and hexadecapole harmonics
\cite{H92}, \cite{CFW94}
\begin{equation} 
 \sP ( {\sk} ) =
   \sP_0 ( k ) {\rm P}_0 ( \mu ) +
   \sP_2 ( k ) {\rm P}_2 ( \mu ) +
   \sP_4 ( k ) {\rm P}_4 ( \mu )
\end{equation}
(${\rm P}_l ( \mu )$ here are Legendre polynomials)
with
\begin{equation} 
\label{harmonics}
   \sP_0 ( k ) = ( 1 + \frac{2}{3} \ff + \frac{1}{5} \ff^2 ) P ( k )
  \ , \ \ \ 
   \sP_2 ( k ) = ( \frac{4}{3} \ff + \frac{4}{7} \ff^2 ) P ( k )
  \ , \ \ \ 
   \sP_4 ( k ) = \frac{8}{35} \ff^2 P ( k )
  \ .
\end{equation}
All the higher order harmonics are predicted to be zero in the linear regime
(odd harmonics vanish in any case by pair exchange symmetry,
while $m \neq 0$ azimuthal harmonics vanish by symmetry
about the line of sight).
The hexadecapole harmonic $\sP_4 ( k )$
is predicted by equation~(\ref{harmonics}) also to be quite small,
only 1/10th of the amplitude of the monopole and quadrupole harmonics
even for $\ff = 1$,
so in practice almost all the information about redshift distortions
in the linear regime is contained in the monopole and quadrupole
harmonics $\sP_0 ( k )$ and $\sP_2 ( k )$.
The ratio $\sP_2 ( k ) / \sP_0 ( k )$ of these harmonics
of the redshift power spectrum then provides a way to measure $\ff$,
hence $\Omega$, independently of the shape of the power spectrum
\begin{equation} 
\label{ratio}
 {\sP_2 ( k ) \over \sP_0 ( k )}
   = {\frac{4}{3} \ff + \frac{4}{7} \ff^2 \over
     1 + \frac{2}{3} \ff + \frac{1}{5} \ff^2}
 \ .
\end{equation}

\section{Measuring the Power Spectrum}
In measuring the distortion of the power spectrum in a real redshift survey,
it is essential to disentangle the true distortion
from the artificial distortion introduced by a non-uniform mask
and selection function.
Now in real space the observed galaxy density is the product of the
true density with the catalogue window,
whereas in Fourier space this becomes a convolution.
Thus the natural place to `deconvolve' the observations from the catalogue
window is real space,
where deconvolution reduces to division,
and where the observations exist in the first place.
The procedure is described in some detail in \cite{H93b}.
In brief,
in computing the contribution to the power spectrum
from a pair of galaxies at separation $r$ and angle $\theta$
to the line of sight,
one divides by the probability of finding a pair anywhere in the
catalogue with that separation $r$ and angle $\theta$.
The deconvolution procedure is `exact', stable in practice,
and admits minimum variance pair weighting.

The harmonics $\sP_l$ of the redshift power spectrum are related to those
$\sxi_l$ of the redshift correlation function by
\begin{equation} 
\label{Pl}
 \sP_l ( k ) = i^l
  \int_0^{\infty} \sxi_l ( r ) j_l ( k r ) 4 \pi r^2 d r
\end{equation}
where $j_l$ are spherical Bessel functions.
Just as it is necessary in a real catalogue
to measure the correlation function with finite bins of separation,
so also the power spectrum must be measured with finite resolution
in $k$-space.
The harmonics of the power spectrum, smoothed over some window $W ( k )$,
are related to the harmonics of the correlation function by
\begin{equation}
 \sP_l^{\mbox{\scriptsize smoothed}} ( k ) =
  \int_0^{\infty} \sP_l ( k' ) W ( k' / k ) 4 \pi ( k' / k )^2 d ( k' / k ) =
  \int_0^{\infty} \sxi_l ( r ) W_l ( k r ) 4 \pi r^2 d r
\end{equation}
where the real space window $W_l ( r )$ for the $l$'th harmonic is,
from equation~(\ref{Pl}),
the smoothed spherical Bessel function
\begin{equation} 
 W_l ( r ) =  i^l
  \int_0^{\infty} W ( k ) j_l ( k r ) 4 \pi k^2 d k
 \ .   
\end{equation}
I choose to use a smoothing window of the form
\begin{equation}
\label{W}
  W ( k ) \propto k^{n} e^{- k^2/4}
\end{equation}
which has the desirable properties of being positive definite in $k$-space,
and yielding gaussian convergence in real space for harmonics $l \leq n$
provided $n$ is an even integer.
The real space windows corresponding to the window (\ref{W}) are
(if $W ( k )$ is normalised so $\int W ( k ) 4 \pi k^2 d k = 1$)
\begin{equation}
\label{Wl}
  W_l ( r ) =
  {i^l [(n\!-\!l)/2]! \over (3/2)_{(n/2)}} \:
  r^l e^{-r^2} L_{(n-l)/2}^{l\,+\,1/2} ( r^2 )
\end{equation}
where $L_{\nu}^{\lambda}$ are Laguerre polynomials,
and
$(3/2)_{(n/2)} = \Gamma [(n\!+\!3)/2] / \Gamma (3/2)$
is a Pochhammer symbol.
These real space windows (\ref{Wl})
bear an amusing resemblance to hydrogenic wave functions.
The resemblance extends to the quantisation condition
that the index $( n\!-\!l ) /2$ must be a nonnegative integer,
to ensure gaussian convergence at large separations.

Computation of the smoothed harmonics $\sP_l ( k )$
of the redshift power spectrum thus
reduces to taking a suitably weighted sum over galaxy pairs of
$W_l ( k r )$ times ${\rm P}_l ( \cos \theta )$,
the $l$'th Legendre polynomial.
No correction factors for the shape of the power spectrum are necessary
in measuring $\ff$ from the ratio (\ref{ratio})
of the smoothed quadrupole and monopole harmonics of the power spectrum.

The power spectra shown in this paper are smoothed with the window (\ref{W})
with $n = 2$, the smallest $n$ which allows the quadrupole power
to be measured.
This smallest $n$ yields the broadest smoothing window
of the form (\ref{W}),
so the errors in the ratio $\sP_2 ( k )/\sP_0 ( k )$
of smoothed power spectrum harmonics are smallest for this $n$,
though the results at different $k$ are also then most correlated.

\section{Result for $\ff$}

\begin{figure}
\begin{center}
\leavevmode
\epsfbox{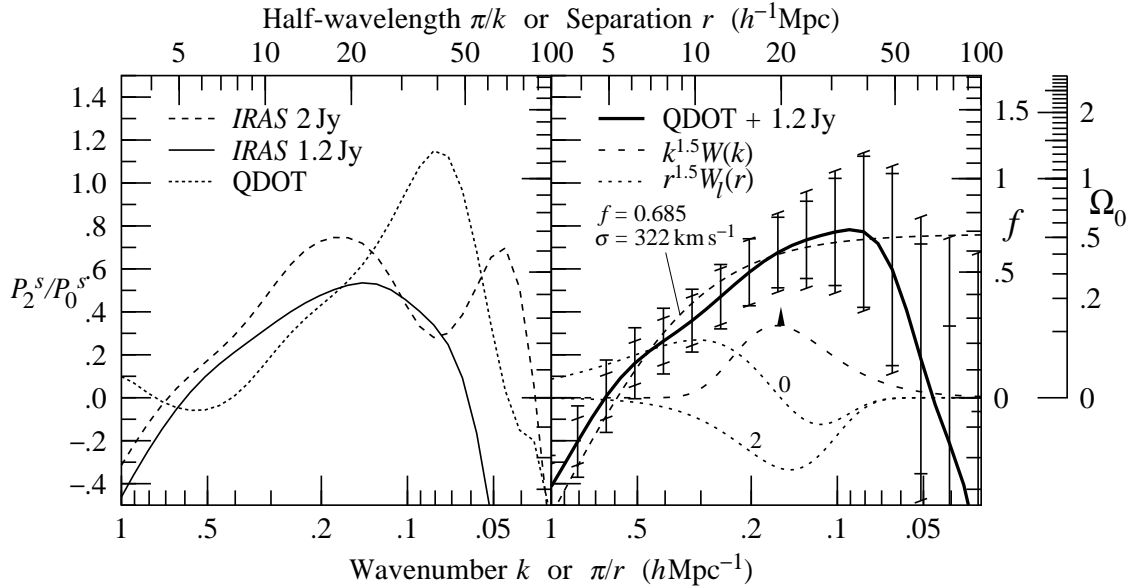}
\end{center}
\caption{
\label{P2P0sum}
\baselineskip 0.4cm 
\protect\small
Ratio of the quadrupole to monopole redshift power spectrum
in the {\em IRAS\/} surveys.
The left panel shows separately
the 2~Jy, 1.2~Jy, and QDOT surveys,
while the right panel shows QDOT plus 1.2~Jy merged
over the angular region of the sky common to both surveys.
Horizontal and tilted bars on the $1 \sigma$ errors
on the QDOT + 1.2~Jy data
distinguish two methods for estimating the uncertainties,
described in \S\protect\ref{error}.
The dashed curve is a fit with $\ff = 0.685$ and
a 1-dimensional pairwise exponential velocity dispersion of $322 \: \kms$.
The smoothing window $W ( k )$, equation~(\protect\ref{W}) with $n = 2$,
of constant width in $\log k$,
gives an impression of the covariance between plotted points.
Notice that the smoothing window is broad,
so the points are correlated.
Also shown are the equivalent real space smoothing windows $W_l ( r )$,
equation~(\protect\ref{Wl}),
for the monopole and quadrupole harmonics of the correlation function.
}
\end{figure}

Figure~\ref{P2P0sum} shows the ratio $\sP_2/\sP_0$
of the quadrupole to monopole power spectra in the {\em IRAS\/}
2~Jy, 1.2~Jy, and QDOT surveys,
and also in QDOT plus 1.2~Jy
merged over the angular region of the sky common to both surveys.
At small scales, $\pi / k < 5 h^{-1} {\rm Mpc}$,
the quadrupole power is negative, which is the nonlinear finger-of-god effect,
while at larger scales the quadrupole power is positive,
which is the squashing effect.

Figure~\ref{P2P0sum} shows a best fit
to the merged QDOT plus 1.2~Jy survey
of a model in which the linear distortion is modulated by
a random exponentially distributed pairwise velocity dispersion,
which seems to offer a simple but adequate description of nonlinearity
\cite{Fisher94}.
The best fitting value of the linear growth parameter $\ff$,
equation (\ref{fb}), is
$\ff = 0.69^{+ .21}_{- .19}$ ($1 \sigma$).

The reader should beware that the measured value of $\ff$
is somewhat sensitive to the relative weight assigned to the near versus far
regions of the survey
(this reflects the problem discussed in \S\ref{systematics};
the above-quoted statistical error on $\ff$
excludes this systematic uncertainty).
Differences between values of $\ff$ obtained by different authors,
\cite{CFW95}, \cite{Fisher94}, \cite{FSL}, \cite{H93a}, \cite{HT},
may be attributable at least in part to differences
in pair weighting schemes.
The turnover in $\sP_2/\sP_0$ evident in Figure~\ref{P2P0sum}
at the largest scales, $\pi / k \gsim \; 50 h^{-1} {\rm Mpc}$,
may be caused in part by the same problem,
although here the uncertainties become large.

\section{Error Estimation}
\label{error}
Uncertainties are estimated by two distinct methods,
distinguished in Figure~\ref{P2P0sum} (\& \ref{P2P0})
by horizontal and tilted bars on the errors.
Both methods take proper account of the correlated character
of the galaxy distribution.

The first method (horizontal bars)
is described and justified in detail in \cite{H93b}.
The procedure is to subdivide the survey into a few hundred volume elements,
measure the fluctuation in the power attributable to each volume element,
and form a variance by adding up the covariances of the fluctuations
between the volume elements.
However, there is an integral constraint which implies that
the sum of all covariances between all volume elements is exactly zero,
so it is necessary to truncate the sum at some point.
The procedure adopted is to order the covariances in order of increasing
distance between each pair of volume elements,
and take the variance to be the maximum value of the cumulative covariance.

The advantage of this method
is that it makes minimal assumptions about the origin and nature of the errors.
Moreover,
systematic errors may be discernible
as systematic patterns in the distribution of fluctuations.
The main defect of the method is that it necessarily underestimates
the variance on scales approaching the scale of the catalogue.

The second method (tilted bars),
inspired by \cite{FKP},
is to take the Poisson (i.e.\ self-pair) variance weighted by
$[ 1 + \Phi_i P ( k ) ] [ 1 + \Phi_j P ( k ) ]$
where $\Phi_i$, $\Phi_j$ are the selection functions at the positions
of a pair $ij$.
The method has the defect that it is properly valid only
for gaussian fluctuations,
but it should be correct to a factor of order unity
also for nongaussian fluctuations.

\section{Systematics}
\label{systematics}
Ideally one would like to `map out' the redshift distortions,
to check that the signal is generally consistent through the survey,
and is not dominated by just a few structures.

\begin{figure}
\begin{center}
\leavevmode
\epsfbox{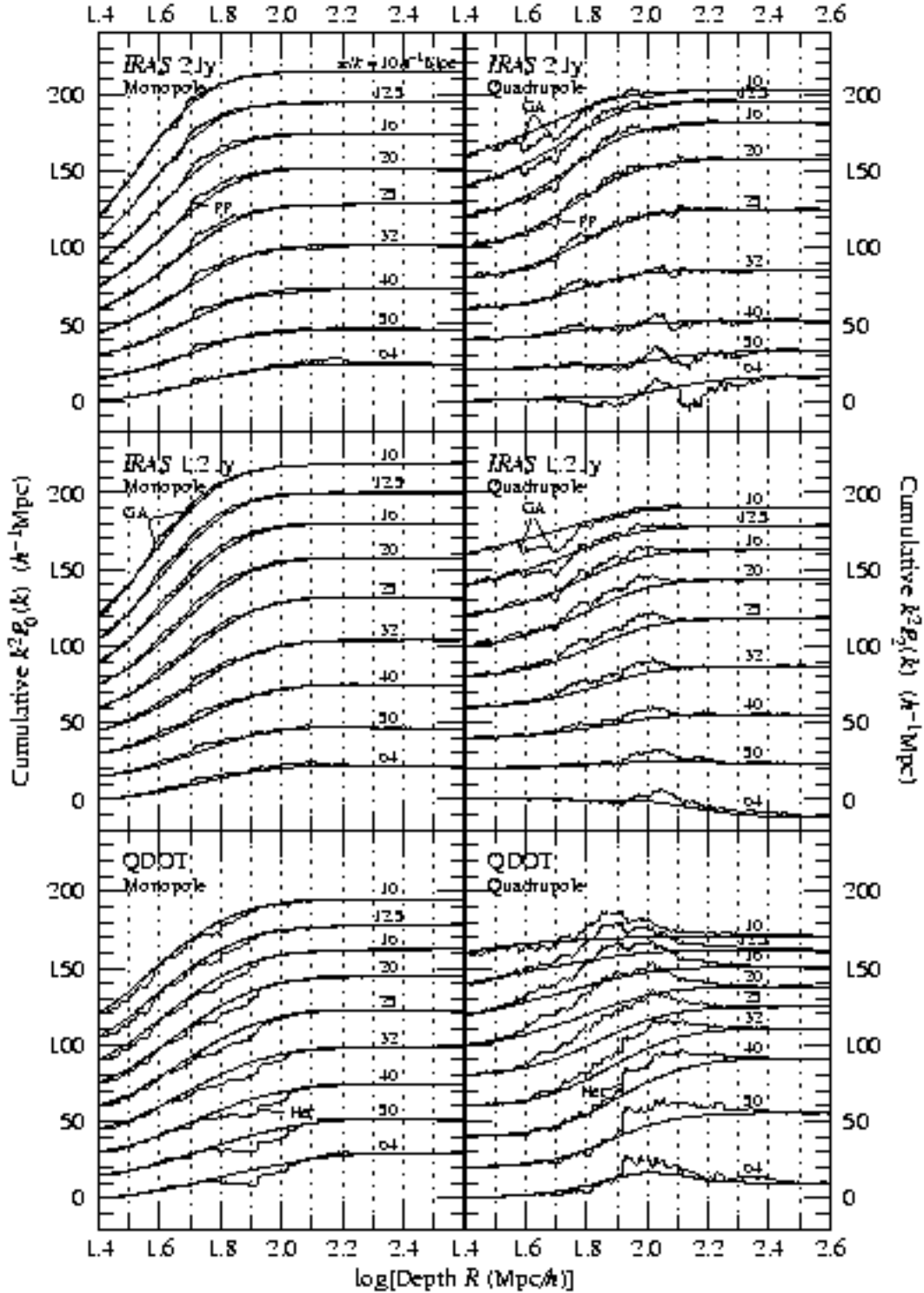}
\end{center}
\caption{
\label{KS}
\baselineskip 0.4cm 
\protect\small
Cumulative monopole (left) and quadrupole (right) power
contributed by many volume elements in the {\em IRAS} 2~Jy (top),
1.2~Jy (middle), and QDOT (bottom) surveys.
There are nine curves in each case,
each corresponding to a different fixed $k$,
with half-wavelength $\pi / k$ as labelled.
The horizontal axis cycles through 22 angular regions,
keyed in Figure~\protect\ref{maps},
within each of 12 radial shells of width 0.1~dex,
from $10^{1.4} h^{-1} {\rm Mpc} \approx 25 h^{-1} {\rm Mpc}$
to $10^{2.6} h^{-1} {\rm Mpc} \approx 400 h^{-1} {\rm Mpc}$,
although in practice there is little contribution to power
from the outermost shells (thanks to the minimum variance pair weighting).
Along with each observed `bumpy' curve
is a smooth curve which shows how the cumulative power
would increase if the clustering around each volume element
were the same as for the full survey.
The curves are offset to separate them on the graph,
and scaled by $k^2$ so that the curves at different $k$
vary with similar amplitude.
}
\end{figure}

\begin{figure}
\begin{center}
\leavevmode
\epsfbox{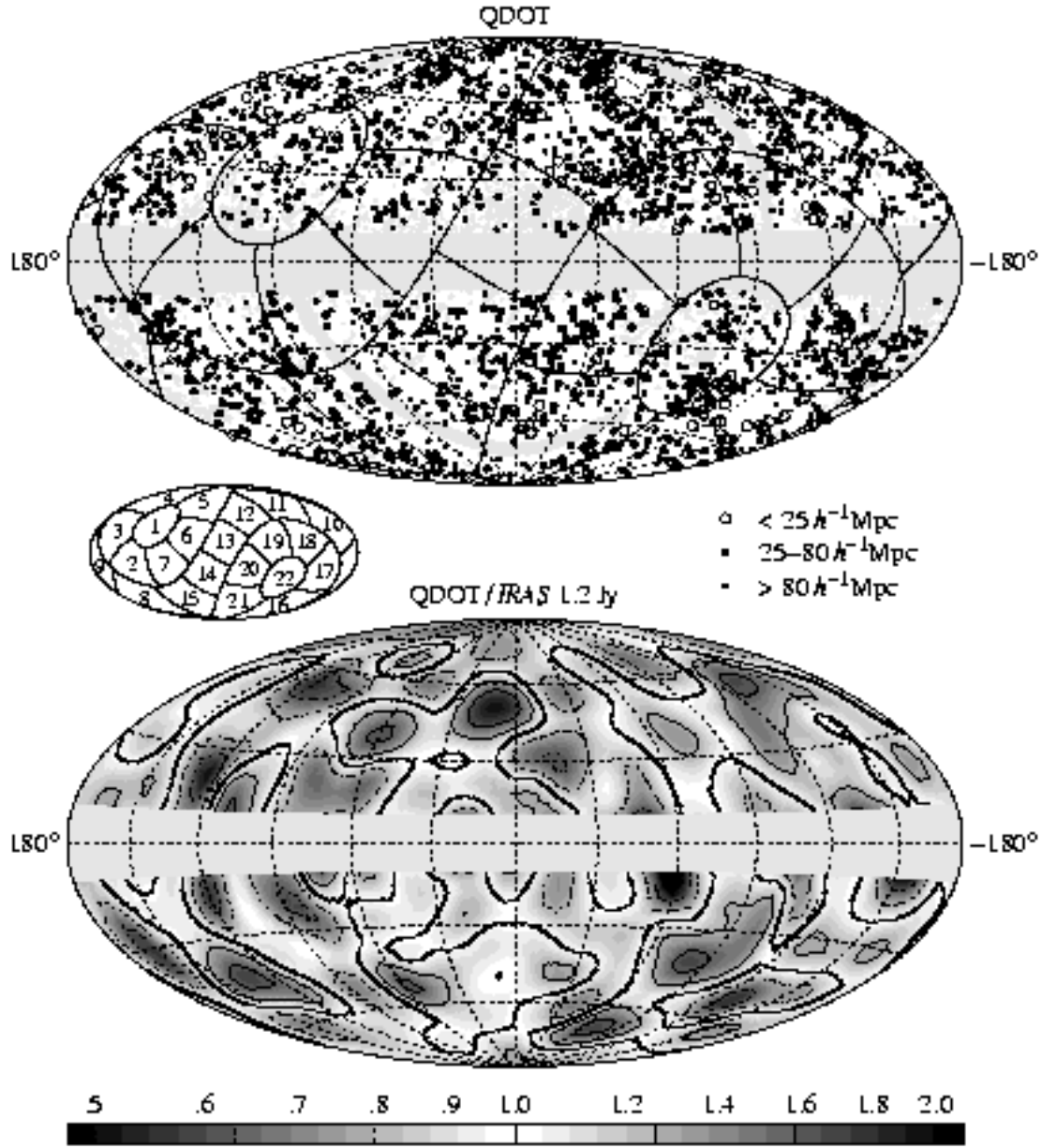}
\end{center}
\caption{
\label{maps}
\baselineskip 0.4cm 
\protect\small
Top:
Map of the QDOT galaxy distribution in galactic coordinates.
Inset:
The 22 angular regions used in the error analysis
and in the `Kolmogorov-Smirnov' plots in Figure~\protect\ref{KS}.
The boundaries of the angular regions are lines of constant ecliptic
latitude and longitude; angular region 1 is the North ecliptic cap.
Bottom:
Ratio of QDOT to 1.2~Jy density on the sky.
The 1.2~Jy density was radially weighted
by the ratio of radial selection functions
to give the same density run with depth as QDOT.
Only galaxies in the area common to both surveys were used,
and the local region within $25 h^{-1} {\rm Mpc}$ was excluded.
The galaxy densities were smoothed with a gaussian with a $1\sigma$
width of $7^\circ$ before taking their ratio.
The thick contour signifies a ratio of unity;
other contours are at intervals of 0.1~dex,
solid for positive fluctuations, broken for negative fluctuations.
Shading is graduated at intervals of 0.02~dex.
}
\end{figure}

Figure~\ref{KS} is an attempt to present such a detailed breakdown
of the distortion signal,
in what can be thought of as `Kolmogorov-Smirnov' type plots.
The error analysis (method one) described in \S \ref{error}
provides a list of fluctuations of the power
in many volume elements in the catalogue.
Figure~\ref{KS} plots these fluctuations cumulatively
as a function of depth, and within each shell of depth
as a function of 22 angular regions on the sky (Figure~\ref{maps}, inset).
In interpreting the plots in Figure~\ref{KS}
one should bear in mind that the fluctuations are correlated.

\begin{figure}
\begin{center}
\leavevmode
\epsfbox{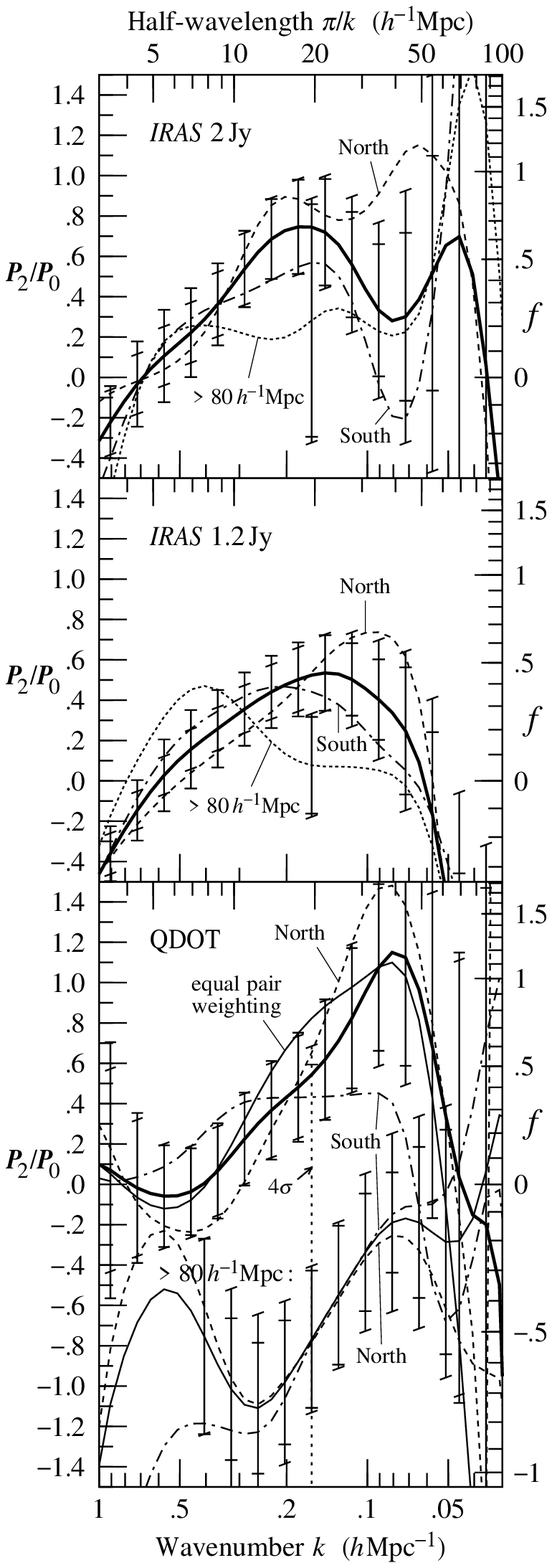}
\end{center}
\caption{
\label{P2P0}
\baselineskip 0.4cm 
\protect\small
Ratio of the quadrupole to monopole power spectrum
in the {\em IRAS\/} 2~Jy, 1.2~Jy, and QDOT surveys,
showing separately the North and South galactic hemispheres,
and the region beyond $80 h^{-1} {\rm Mpc}$.
Also shown in QDOT is
the result of applying equal pair weighting
(which gives relatively more weight to the nearer region than
minimum variance pair weighting).
The smoothing window is the same as plotted in
Figure~\protect\ref{P2P0sum}.
Estimation of $1 \sigma$ errors is described in \S\protect\ref{error}.
}
\end{figure}

The cumulative distribution of power in the 2~Jy survey
(Figure~\ref{KS}, top) looks encouraging.
The most prominent features in the survey are the Great Attractor (GA),
and Perseus-Pisces (PP) complexes \cite{S92a}.
The GA, which is elongated along the line of sight,
produces the negative steps in quadrupole power in angular region 19
in the two radial shells at depths $25-40 h^{-1} {\rm Mpc}$,
while PP, which is flattened,
causes the positive step in quadrupole power in angular region 2
at depth $40-50 h^{-1} {\rm Mpc}$.
The overall impression, however,
is that the cumulative quadrupole power rises consistently,
indicating that the squashing effect is pervasive throughout the survey.

By contrast, the cumulative distribution of power in the QDOT survey
(Figure~\ref{KS}, bottom)
looks quite peculiar.
While the quadrupole power increases consistently to a depth of about
$80 h^{-1} {\rm Mpc}$,
beyond this point it turns over and declines, again consistently,
at all measured wavelengths.
Thus while the near region shows the expected squashing effect,
the far region is `finger-of-goddy'.
Associated with the turnover in quadrupole power
is an enhancement, albeit less marked, in monopole power.
Figure~\ref{KS} indicates that at larger wavelengths,
$\pi / k \gsim \; 40 h^{-1} {\rm Mpc}$,
there is a dominant feature in Hercules,
angular region 5 at $80 - 100 h^{-1} {\rm Mpc}$.
However, it will be seen below that Hercules is not the only
problem area.

The 1.2~Jy survey
(Figure~\ref{KS}, middle)
appears intermediate between the 2~Jy and QDOT surveys.

The right panels of Figure~\ref{xicont}, which show
contour plots of the redshift correlation function in the region
beyond $80 h^{-1} {\rm Mpc}$ depth,
confirm visually the indications from Figure~\ref{KS}.
For QDOT,
the correlation function of the far region
(Figure~\ref{xicont}, bottom right)
shows a broad extension along the line of sight,
to separations of $50 h^{-1} {\rm Mpc}$ and more,
in contrast to the squashing effect observed in the full survey
(Figure~\ref{xicont}, bottom left).
There is a sign of a similar, albeit milder, extension in the
far region of the 1.2~Jy survey
(Figure~\ref{xicont}, middle right).
There may even be some hint of the effect in 2~Jy
(Figure~\ref{xicont}, top right),
although the noise here is too great to draw any conclusion.

Figure~\ref{P2P0} provides a quantitative assessment of the statistical
significance of the discrepancy between the near and far regions.
It shows the ratio $\sP_2/\sP_0$ of the quadrupole to monopole power
in each of the 2~Jy, 1.2~Jy and QDOT surveys,
showing separately the region beyond $80 h^{-1} {\rm Mpc}$,
and also the North and South galactic hemispheres.
The first point to notice is that the North and the South
are consistent with each other in all cases,
fluctuating from the total at around the $1\sigma$ level,
as expected for two almost independent halves of a sample.

In the 2~Jy (Figure~\ref{P2P0}, top) and 1.2~Jy (Figure~\ref{P2P0}, middle) 
surveys the $\sP_2/\sP_0$ ratio in
the region beyond $80 h^{-1} {\rm Mpc}$
is again consistent with the full survey,
though there is some hint of a downward fluctuation.
In QDOT however (Figure~\ref{P2P0}, bottom) the ratio
in the region beyond $80 h^{-1} {\rm Mpc}$ is decidedly low
compared to the full survey,
by about $4 \sigma$
at the most accurately measured scale,
$\pi / k \sim 20 h^{-1} {\rm Mpc}$.
The discrepancy in QDOT appears about equally
in both North and South,
demonstrating that the problem is not confined to Hercules.
The region beyond $80 h^{-1} {\rm Mpc}$ in QDOT
contains 1071 galaxies,
about half the galaxies in the survey.

\section{Variable Depth of PSC Fluxes over the Sky?}
While the discrepancy between the near and far regions of QDOT
is not so large as to be conclusive,
it is large enough to be worrying.
Certainly it is large enough to influence the measurement
of cosmological quantities.
If the discrepancy is real,
what could be causing it?

Several evidences argue against the problem being errors in redshifts.
In the first place,
if the `finger-of-goddiness' in the far region of QDOT were caused
by redshift errors,
then it would tend to reduce the monopole power in the far region,
whereas the opposite, an enhancement of power, is observed.
Secondly,
the `finger-of-god' in the far region extends to $5{,}000 \: \kms$ and more,
which seems implausibly large for redshift errors.
Thirdly,
tests of QDOT redshifts against others show no large systematic differences.

The effect
appears about equally in intrinsically faint and luminous galaxies.
This argues against the hypothesis that the `finger-of-goddiness'
of the far region results from
the tendency of infrared-luminous galaxies to occur in interacting systems.

Other tests
show the effect to be insensitive to details of the analysis.
For example,
even gross changes in the radial selection function
have surprisingly little effect.
The effect is not caused by cosmological evolution,
whether in the volume element, in fluxes, or in the correlation function.

A possible explanation of the problem is that there is some
systematic variation in the effective depth of the {\em IRAS\/}
Point Source Catalog (PSC) across the sky.
Such an explanation would be consistent with several facts:
(1)
Systematic variations in flux limit over the sky would both
increase the overall (monopole) power,
and at the same time cause more deeply sampled regions to appear
elongated along the line of sight.
This agrees with what is observed.
(2)
The change in the pattern of clustering appears at a depth
of about $80 h^{-1} {\rm Mpc}$,
which is where the selection function in QDOT starts to steepen.
Flux variations will cause larger variations in density where
the selection function is steeper.
(3)
The effect is most marked in QDOT,
whose flux limit of 0.6~Jy is close to the 0.5~Jy limit of the PSC.

The effect is probably {\em not\/} caused by real intrinsic variations
in the luminosity function of different regions,
because in that case the 2~Jy and 1.2~Jy surveys would be as affected
as QDOT.

\begin{figure}
\begin{center}
\leavevmode
\epsfbox{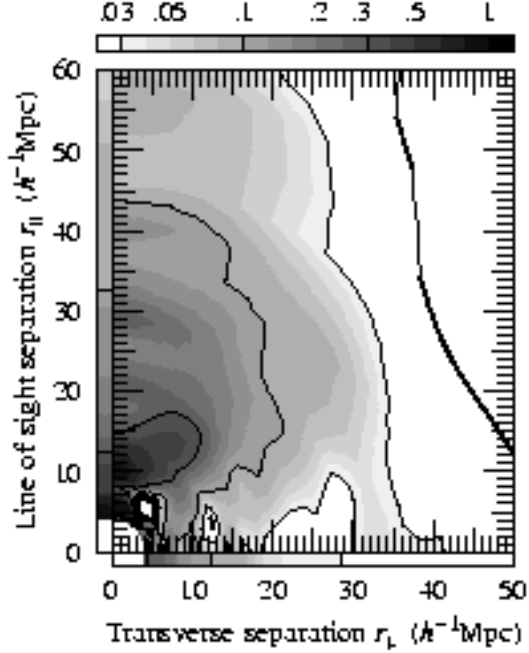}
\end{center}
\caption{
\label{xiw}
\baselineskip 0.4cm 
\protect\small
Contour plot of $(\sxi_{> 80} - \sxi) / ( 1 + \sxi )$ in QDOT,
where $\sxi$ and $\sxi_{> 80}$ denote the redshift correlation functions
respectively of the full survey and of the region beyond $80 h^{-1} {\rm Mpc}$,
as plotted in Figure~\protect\ref{xicont}, bottom.
The plotted quantity represents the correlation
function $\sxi_{\phi}$ of a variable angular selection function $\phi$,
equation~(\protect\ref{xiphi}),
which hypothetically modulates the density in the far region of QDOT.
Contour levels are the same
as in Figure~\protect\ref{xicont}.
Bars along the two axes are contour plots of the $1\sigma$
uncertainties in the parallel and perpendicular directions,
as in Figure~\protect\ref{xicont}.
}
\end{figure}

Figure~\ref{xiw} offers a quantitative assessment of the hypothesis of
variations in the flux limit of the PSC across the sky.
The plot is motivated by the following rough mathematical argument.
Suppose flux variations cause a variable selection function $\phi$
across the sky.
The observed galaxy density $\rho_{\rm obs}$ is the product of the true
galaxy density $\rho$ times the variable selection function $\phi$
\begin{equation}
 \rho_{\rm obs} = \rho \phi
 \ .
\end{equation}
The observed correlation function $\sxi_{\rm obs}$ is then related
to the true correlation function $\sxi$ and the correlation function
$\sxi_{\phi}$ of the variable angular selection function by
\begin{equation}
\label{xiobs}
 1 + \sxi_{\rm obs} =
  \langle \rho_{{\rm obs} , 1} \rho_{{\rm obs} , 2} \rangle =
  \langle \rho_1 \phi_1 \rho_2 \phi_2 \rangle =
  \langle \rho_1 \rho_2 \rangle \langle \phi_1 \phi_2 \rangle =
  ( 1 + \sxi ) ( 1 + \sxi_{\phi} )
 \ .
\end{equation}
The third equality of equation~(\ref{xiobs}) assumes that
the true density $\rho$ and the selection function $\phi$ are uncorrelated,
which may be false.
Equation (\ref{xiobs}) rearranges into an expression for the correlation
function $\sxi_{\phi}$ of the variable angular selection function,
\begin{equation}
\label{xiphi}
 \sxi_{\phi} =
  {\sxi_{\rm obs} - \sxi \over 1 + \sxi}
 \ .
\end{equation}

Now take the `observed' correlation function $\sxi_{\rm obs}$
in equation~(\ref{xiphi})
to be the correlation function
$\sxi_{> 80}$
of QDOT beyond $80 h^{-1} {\rm Mpc}$
(which is more affected by flux variations),
and take the `true' correlation function $\sxi$ to be the correlation
function of the full QDOT sample
(which is less affected).
If the hypothesis of a variable flux limit across the sky is true,
then Figure~\ref{xiw} represents the correlation function
$\sxi_{\phi}$ of the resulting variable angular selection function
(more precisely, the difference in the angular selection functions
in the $> 80 h^{-1} {\rm Mpc}$ versus full samples).

The prediction then is that the contours in Figure~\ref{xiw},
which by hypothesis represents the correlation function $\sxi_{\phi}$
of an angular selection function,
should be more or less vertical
(not quite vertical because the horizontal axis is a physical rather than an
angular separation, but this should not make too much difference
at a depth of $80 h^{-1} {\rm Mpc}$).
Now certainly there are structures in the contours of Figure~\ref{xiw}
(for example, the blob on the line of sight axis at
$r_\parallel \approx 10 h^{-1} {\rm Mpc}$
{\em is\/} produced by an excess of luminous interacting pairs),
but the general impression is that the contours are consistent with
being more or less vertical.

Suppose that the $\sxi_{\phi} = 0.1$
contour level in Figure~\ref{xiw} is believable.
The implied rms variation in the density is
$\sxi_{\phi} ^{1/2} = 0.1^{1/2} \approx 0.3$.
The density variation can be converted into a flux variation
by dividing by
the rms value of the logarithmic slope $d \ln \Phi / d \ln R^2$
of the radial selection function $\Phi$ with respect to depth $R$.
This rms slope is
$\langle ( d \ln \Phi / d \ln R^2 )^2 \rangle^{1/2} \approx 1.7$
(the Euclidean value would be $3/2$)
for QDOT beyond $80 h^{-1} {\rm Mpc}$,
implying an rms flux variation of $0.3 / 1.7 \approx 0.18$.
At the 0.6~Jy limit of the QDOT,
this translates into a flux variation of
$0.18 \times 0.6 \: {\rm Jy} \approx 0.10$~Jy.
Figure~\ref{xiw} suggests that the variation increases somewhat to
smaller scales;
the scale at which the variation is $0.10$~Jy
appears to be $r_\perp \sim 15 h^{-1} {\rm Mpc}$,
corresponding to an angular scale of $\sim 7^\circ$.

A failing of the above argument is that flux variations should
cause at least some line-of-sight elongation of the correlation function
also in the near region of QDOT,
whereas no obvious elongation is observed.
The amplitude of the effect in the near region is reduced by
two effects, the shallower slope of the selection function,
and the fact that a given transverse separation $r_{\perp}$
corresponds to a larger angular separation, where the flux variations
are (apparently) smaller.
All the same,
the absence of the effect in the near region is surprising.

\section{Map}
\label{map}

If there are systematic flux variations in the PSC which affect QDOT
more than 1.2~Jy,
then one way to try and map out the hypothesised flux variations on the sky
is to take the ratio of the QDOT and 1.2~Jy densities on the sky.
Figure~\ref{maps} (bottom) shows the result.
The most prominent overdensities in QDOT relative to 1.2~Jy
are Hercules
($l = 40^\circ$, $b = 40^\circ$; the peak at $l = 10^\circ$, $b = 50^\circ$
is actually in Serpens)
and Leo
($l = 230^\circ$, $b = 50^\circ$)
in the North,
and Pisces
($l = 130^\circ$, $b = -40^\circ$)
and Reticulum
($l = 280^\circ$, $b = -40^\circ$)
to Phoenix
($l = 290^\circ$, $b = -70^\circ$)
in the South.
There is some indication that regions close to the Galactic plane
are underdense in QDOT relative to 1.2~Jy,
though bear in mind that these and other heavily masked regions are noiser:
the Great Attractor region ($l = 310^\circ$),
Cygnus ($l = 80^\circ$),
Cassiopeia ($l = 130^\circ$),
and Taurus ($l = 170^\circ$, $b = - 30^\circ$).
Other underdense regions include
Draco
($l = 110^\circ$, $b = 50^\circ$)
and Eridanus
($l = 230^\circ$, $b = - 50^\circ$).
There is some indication that the 2HCON regions
are underdense in QDOT relative to 1.2~Jy
(these are two D-shaped regions to the ecliptic east (left and down)
of the two masked strips in the top map of Figure~\ref{maps};
see \cite{IRAS}~Figure~III.D.3).

While Figure~\ref{maps} may contain some clues as to the distribution of
hypothetical flux variations,
such inferences should be viewed skeptically,
for the level of fluctuations in the map is in fact entirely consistent
with Poisson noise fluctuations between QDOT and 1.2~Jy.

\section{Conclusion}
The large scale squashing effect predicted by linear theory
is clearly observed in the {\em IRAS} 2~Jy, 1.2~Jy, and QDOT surveys.
However,
the pattern of redshift distortions changes markedly in QDOT
beyond its median depth of $80 h^{-1} {\rm Mpc}$,
becoming extended along the line of sight rather than squashed.
There is some sign of a similar effect at a mild level in the 1.2~Jy survey.
I have argued that systematic variations in the depth of the
{\em IRAS} PSC fluxes of $\sim 0.1$~Jy on scales $\sim 7^\circ$
could account for several aspects of this discrepancy.

\acknowledgements{I thank George Efstathiou for the hospitality
of the Nuclear \& Astrophysics Laboratory, Oxford University,
where this paper was written.
This work was supported by a PPARC Visiting Fellowship,
NSF grant AST93-19977,
and NASA grant NAG 5-2797.}

\vfill

\begin{thebibliography}{99}{\baselineskip 0.4cm
\bibitem{IRAS} Beichman C.\ A., Neugebauer G., Habing H.\ J., Clegg P.\ E., Chester T.\ J., 1988, {\it IRAS Catalogs Volume 1 Explanatory Supplement}, NASA
\bibitem{CFW94} Cole S., Fisher K.\ B., Weinberg D.\ H., 1994, \mnras {267} {785}
\bibitem{CFW95} Cole S., Fisher K.\ B., Weinberg D.\ H., 1995, {\em Mon.\ Not.\ R.\ Astr.\ Soc.\/} submitted
\bibitem{FKP} Feldman H.\ A., Kaiser N., Peacock J.\ A., 1994, \apj {426} {23}
\bibitem{Fisher94} Fisher K.\ B., Davis M., Strauss M.\ A., Yahil A., Huchra J.\ P., 1994, \mnras {267} {927}
\bibitem{12Jy} Fisher K., Huchra J., Strauss M., Davis M., Schlegel D., 1995, {\em Astrophys.\ J.\ Suppl.\ Ser.\/} in press
\bibitem{FSL} Fisher K.\ B., Scharf C.\ A., Lahav O., 1994, \mnras {266} {219}
\bibitem{Kaiser} Kaiser N., 1987, \mnras {227} {1}
\bibitem{H92} Hamilton A.\ J.\ S., 1992, \apj {385} {L5}
\bibitem{H93a} Hamilton A.\ J.\ S., 1993, \apj {406} {L47}
\bibitem{H93b} Hamilton A.\ J.\ S., 1993, \apj {417} {19}
\bibitem{HT} Heavens A.\ F., Taylor A.\ N., 1995, {\em Mon.\ Not.\ R.\ Astr.\ Soc.\/} in press
\bibitem{Lawrence} Lawrence A., Rowan-Robinson M., Crawford J., Parry I., Xia X.-Y., Ellis R.\ S., Frenk C.\ S., Saunders W., Efstathiou G., Kaiser N., 1995 in preparation (QDOT)
\bibitem{S90} Strauss M.\ A., Davis M., Yahil A., Huchra J.\ P., 1990, \apj {361} {49}
\bibitem{S92a} Strauss M.\ A., Davis M., Yahil A., Huchra J.\ P., 1992, \apj {385} {421}
\bibitem{S92b} Strauss M.\ A., Huchra J.\ P., Davis M., Yahil A., Fisher K.\ B., Tonry J., 1992, \apjs {83} {29}
}
\end{thebibliography}
\end{document}